\documentclass [12pt,a4paper,draft]{article}

\usepackage{times}

\DeclareFontFamily{OT1}{times}{}
\DeclareFontShape {OT1}{times}{m }{n }{ <-> ptmr }{}
\DeclareFontShape {OT1}{times}{bx}{n }{ <-> ptmb }{}
\DeclareFontShape {OT1}{times}{m }{it}{ <-> ptmri}{}
\DeclareFontShape {OT1}{times}{bx}{it}{ <-> ptmbi}{}
\usepackage{amsmath}
\usepackage{amsfonts}
\usepackage{amssymb}
\usepackage{latexsym}
\begin{document}

\title{\vspace{-3.cm} \bf A comparison of delayed radiobiological effects of depleted--uranium munitions versus fourth--generation nuclear weapons}
\author{Andr\'e Gsponer, Jean-Pierre Hurni, and Bruno Vitale\\
\emph{Independent Scientific Research Institute}\\ 
\emph{ Box 30, CH-1211 Geneva-12, Switzerland}\\
e-mail: isri@vtx.ch\\ 
~ \\ 
{\bf Contributed to \emph{YUNSC--2002}}\\
Belgrade, Yugoslavia, Sep. 30 -- Oct. 4, 2002}

\date{10 October 2002}

\maketitle

\begin{abstract}

It is shown that the radiological burden due to the battle-field use of circa 400 tons of depleted-uranium munitions in Iraq (and of about 40 tons in Yugoslavia) is comparable to that arising from the hypothetical battle-field use of more than 600 kt (respectively 60 kt) of high-explosive equivalent pure-fusion fourth-generation nuclear weapons.  

Despite the limited knowledge openly available on existing and future nuclear weapons, there is sufficient published information on their physical principles and radiological effects to make such a comparison.  In fact, it is shown that this comparison can be made with very simple and convincing arguments so that the main technical conclusions of the paper are undisputable --- although it would be worthwhile to supplement the hand calculations presented in the paper by more detailed computer simulations in order to consolidate the conclusions and refute any possible objections.

From a strategic perspective, the breaking of the taboo against the intentional battle-field use of radioactive materials, which lasted from 1945 to 1991, can therefore be interpreted as a preparation for the progressive introduction of fourth-generation nuclear weapons whose battle-field use will cause a low (but non-negligible) residual radioactive environment.  It can therefore be argued that besides its military function, the use of depleted-uranium in Iraq and Yugoslavia may have served a political purpose: to soften the opposition of the Western public opinion to the induction of radioactivity on the battle-field, and to get the World population accustomed to the combat use of depleted-uranium and fourth-generation nuclear weapons.

\end{abstract}

\section{Introduction}

The last two major military campaigns in which the full extent of the Western military superiority was displayed --- Iraq and Yugoslavia --- were characterized by the extensive use of precision guided delivery systems and  depleted-uranium munitions \cite{GUTVI2002-}.\footnote{According to official accounts, no depleted-uranium ammunition was used in the more recent war in Afghanistan.}

While there has been some political opposition in a number of countries to the battle-field use of depleted-uranium projectiles, and considerable reaction from many environmental organizations world-wide, there has been little analysis of the strategic-political reasons why such ammunition was used in the first place.  Of course, it has been stressed that opposition to these weapons could have been avoided if the low-radioactive U-238 would have been replaced by some non-radioactive material such as tungsten.  However, the possible link of the use of depleted-uranium to a gradual shift towards the actual battle-field use of nuclear weapons seems not to have been fully appreciated.

The purpose of this paper is to show that despite the relatively ``low'' radioactive nature of U-238, the radiological impact of its use in conventional ammunition is comparable to that of the battle-field use of hypothetical fourth-generation nuclear weapons in which the explosive power would derive from thermonuclear fusion rather than nuclear fission --- and in which essentially zero fission reactions would take place because the ``fission-trigger'' used in ordinary hydrogen bombs would be replaced by some other device \cite{GSPON2000-}.

Fourth-generation nuclear weapons are under active development in all major nuclear weapon states --- and all major industrial powers such as Germany and Japan are actively developing the technology required to develop such weapons.  In a nut-shell, the principle of these new types of nuclear weapons is to trigger (by some advance technology such as a superlaser, magnetic compression, antimatter, etc.) a relatively small thermonuclear explosion in which a deuterium-tritium mixture is burnt in a device whose weight and size are not much larger than a few kilograms and liters.  Since the yield of these warheads could go from a fraction of a ton to several tens, or hundreds, of tons of high-explosive equivalent, their delivery by precision guided munitions or other means will dramatically increase the fire-power of those who possess them --- without crossing the threshold of using kiloton to megaton size nuclear weapons, and therefore without breaking the taboo against the first-use of weapons of mass-destruction.

Consequently, if the precision guided delivery systems used in Iraq and Yugoslavia would have been loaded with fourth-generations nuclear warheads rather than conventional explosives, much less of them could have been used to achieve the same military objective --- although at the cost of a radiological burden that we estimate in this paper and compare to that of depleted-uranium munitions.

The somewhat surprising fact that the medium- to long-term radiological burden arising from the large-scale battle-field use of depleted-uranium is comparable to that of many kilotons equivalent of fourth-generation nuclear weapons stems from the fact that in these weapons the main source of explosive power is fusion rather than fission.  Therefore, the radiological burden induced by the detonation of fourth-generation nuclear weapons derives primarily from the unburnt tritium that is dispersed in the environment by the explosion, and secondarily from the activation of the air, ground, and concrete structures by the neutrons that are produced during the explosion.  Since tritium is a relatively ``low'' radioactive material, very significative amounts of tritium are required to produce a radiological burden comparable to that of the U-238 expended in Iraq or Yugoslavia.  This explains why it is meaningful to compare the radiological effect of the conventional-weapon's use of a relatively harmless nuclear material such as U-238 to the nuclear-weapon's use of the most important thermonuclear fuel: tritium. 

Throughout the paper we tried to remain at the lowest possible technical level in order to make it understandable to as many non-specialists as possible.  Similarly, we cite a variety of references to show that there exist many open sources in which the information used in our paper can be found.  In particular, concerning the technicalities of thermonuclear weapons and their radiological effects, much can be inferred from papers published in the 1950-60s during the debate over the hydrogen bomb and the atmospheric test of nuclear weapons, e.g., \cite{LEIPUN1957-}; papers published during the 1960-70s in support of ``peaceful nuclear explosions,'' e.g., \cite{LESSL1971-,GREEN1972-}; papers published in the 1970-80s during the debate over the ``neutron bomb,'' e.g., \cite{SANDM1972-,SEIFR1984-,SAHIN1985-}; and more recent publications related to the quest for fourth-generation nuclear weapons \cite{GSPON2000-}. 

The organization of this paper is as follows:  In section (1) we calculate an upper limit to the radiological burden due to the battle-field use of depleted-uranium. This burden will be compared in sections 2 and 5 to the corresponding burden calculated by similar methods for the battle-field use of fourth-generation nuclear weapons. In section (2) a simple comparison is made in which the effects of the neutrons produced during the explosion are neglected.  In section (3) we calculate the dominant contributions to the delayed radiological effects of the explosion of pure-fusion fourth-generation nuclear weapons. In section (4) we estimate the magnitude of a fourth-generation nuclear weapon's engagement equivalent to the radiological burden due to the expenditure of a given amount of depleted-uranium ammunition.  Finally, in the conclusion, we summarize and discuss the main points established in the paper.

\section{Upper limit to the radiological burden due to \\the battle-field use of depleted-uranium}

Armour-piercing munitions containing depleted-uranium (DU) were first used during the 1991 Gulf War, therefore breaking a 46 years long taboo against the intentional use or induction of radioactivity in combat.\footnote{Depleted uranium munitions were primarily designed to stop a massive tank attack by the nuclear-armed Warsaw Pact Organization.  Despite of this, when the U.S. military's intention to use uranium bullets became public, the program was denounced in the Congressional Record as ``shocking'' \cite{ARKIN1993-}.} Similar ammunition was used more recently in the conflict involving NATO troops in Yugoslavia.

The DU in these anti-tank weapons is initially in solid form, e.g., in the case of the API PGU 14/B projectiles delivered by A-10 ground-attack aircrafts, in the form of a 96 mm long and 16 mm diameters cylinder which constitutes the main component of the armor-piercing penetrators.

\begin{citation}

   ``When these penetrators strike a hard target, such as a tank, a large fraction of their kinetic energy is converted into heat in less than a millisecond. This rapid release of energy can convert much of the DU into small, hot fragments and particles.  The smaller fragments can burn, generating DU-oxide aerosols.  When a penetrator strikes a soft target, such as a personnel carrier, truck, or soil, much less aerosol is generated and much of the DU penetrator may remain intact'' \cite[p.132]{FETTE1999-}.
\end{citation}

Consequently, there can be much debate about how much of the DU expended in a military campaign has actually been dispersed in the biosphere in the form of aerosols or dust, and how much can be recovered and safely disposed of.

For the purpose of the comparison that is subject of this paper we will make the maximizing assumption that \emph{all} of the DU could have been dispersed so that its radiological impact is directly related to the total amount of DU used in the conflict.  This conservative assumption can of course be criticized on several grounds, and we will come back to it in the concluding section. However, it can already be said that the main conclusion of this study will not be affected by this assumption since a corresponding maximizing assumption can be made concerning the dispersal of radioactive materials in the combat use of fourth-generation nuclear weapons.

The overall radiological burden, which we express in units of ``Roentgen equivalent for men,'' rem,\footnote{We use the ``old'' radiological units (Ci, rad, rem, etc.) in this paper because most of the papers dealing with nuclear weapons, and most of the cited references, still use these units instead of the modern ones: 1 Bq = 1/($3.7 \times 10^{10}$) Ci, 1 Gy = 100 rad, 1 Sv = 100 rem, etc.} is related to the specific radioactivity of the material under consideration, i.e., the number of disintegrations per second and per gram of material.  This specific activity $\alpha$ is a function of the half-life $t_{1/2}$ and the atomic weight $A$, and is conventionally normalized relative to the radioactivity of 1 gram of radium-226 whose half-life is 1620 years.  Therefore,
$$
 \alpha = \frac{1620}{t_{1/2}} \times \frac{226}{A} ~~~ ~~~ \text{[Ci/g]} ~~. \eqno(1)
$$
For depleted-uranium, i.e., U-238, this gives a specific activity of $3.4\times 10^{-7}$ Ci/g which is listed in the last line of Table~1.  In order to translate this activity into a radiological burden assuming that somehow the DU will be intaken by human beings, it is necessary to apply a conversion factor which we borrow from the official regulations based on the recommendations of the \emph{International Commission on Radiation Protection} \cite{OCPCR1976-}. The concept underlying this conversion is that of the ``Equivalent Dose at the Critical Organ'' (EDCO), which results in the conversion factors $D$ listed in Table~1, that gives a specific burden of $\beta_U = 24$~rem per gram of U-238 according to the formula
$$
      \beta = D \times \alpha ~~~ ~~~ \text{[rem/g]}  ~~.  \eqno(2)
$$
The overall burden due to a mass $m_U$ of DU is then
$$
                      B(^{238}U) = \beta_U  m_U ~~~ ~~~ \text{[rem]} \eqno(3)
$$
so that if this equation is applyed to the 400 t of U-238 expended in Iraq, one gets a figure of about $10^{10}$ rem, which constitutes the extreme upper limit to the radiological impact of that amount of DU.

\section{Simplified comparison}

It is well known that in nuclear explosions about 5000 to 20000 Ci/kt of tritium are produced by the fusion reaction, while only about 1 Ci/kt of tritium is produced by the fission process \cite{LESSL1971-}.\footnote{This is why in some applications of ``peaceful nuclear explosions,'' such as gas well stimulation, fission devices surrounded by a neutron shield were preferred to fusion devices, despite the fact that fission devices produce large amounts of fission-products, while fusion devices are comparatively much ``cleaner'' \cite[p.434]{GREEN1972-}}  It is also elementary knowledge that \emph{thermonuclear fusion produces an energy equivalent to 133 t of high-explosive for each gram of tritium consumed}, which corresponds to $1.45 \times 10^{24}$ deuterium-tritium fusion reactions per kt (see, e.g., \cite{GLASS1977-,GSPON2000-}).  One can therefore derive from these figures a simple estimate of the ``tritium-inefficiency'' of a thermonuclear explosive, assuming that all its energy yield comes from the burning of tritium, and that the tritium released in the environment is the amount that has not been burnt during the explosion:
$$
 \epsilon = \frac{\text{tritium not burnt during the explosion}}
             {\text{tritium initially present or produced during the explosion}}
       ~~~. \eqno(4)
$$
This requires to know the specific activity of tritium, which can be calculated using equation (1) and gives the figure listed in the first line of Table~1. Therefore, since  $10^{-4}$ g of tritium corresponds to 1 Ci, the burning of $1/0.0133 \cong 75$ Ci of tritium corresponds to an explosive yield of 1 tons.  This means that according to reference \cite{LESSL1971-} an actual thermonuclear explosive has a tritium-inefficiency of about 7 to 27~\%, which is in fair agreement with reference \cite{LEIPUN1957-} which assumes $\epsilon = 10$~\%.

These estimates show that in terms of tritium economy, thermonuclear explosives are fairly efficient, the efficiency (or burn-up)
$$
 \eta = 1-\epsilon = \frac{\text{tritium burnt during the explosion}}
         {\text{tritium initially present or produced during the explosion}}
       \eqno(5)
$$
being on the order of 73 to 93~\% for the high-yield explosives referred to in references \cite{LEIPUN1957-, LESSL1971-}.

However, for low-yield thermonuclear explosions, whether in the kiloton range as in ``neutron bombs'' \cite{SAHIN1985-},\footnote{The neutron bomb, a typical third-generation nuclear weapon, is not a pure-fusion weapon since about half of its explosive power comes from fission.} or in the sub-kiloton range as in inertial confinement fusion experiments and fourth-generation nuclear weapons, achievement of high efficiencies is much more difficult and an efficiency of 30 to 50 \% is considered as reasonable \cite{GSPON2000-}.  Moreover, in such low-yield devices, much less tritium is bred \emph{in situ} during the explosion so that the efficiency $\eta$ is essentially equal to the amount of tritium burnt over the amount of tritium initially present in the device.

Therefore, if the short term effects and the radioactivity induced by the neutrons produced in the explosion are neglected, one gets a zeroth order approximation of the delayed radiological impact of a fourth-generation nuclear weapon by assuming that the burden is solely due to the dispersal of the unburnt tritium.  Using equations similar to (1) and (2), this burden will be
$$
       B(^3T) = (1 - \eta) \beta_T m_T  ~~, \eqno(6) 
$$
where $\beta_T$ is the specific burden listed on the first line of Table~1, and $m_T$ the mass of tritium initially present in the weapon.

Combining equations (6) and (3) we can therefore estimate the amount of tritium equivalent to a given amount of DU, i.e.,
$$
        m_T = \frac{1}{1 - \eta} \frac{\beta_U}{\beta_T}  m_U  ~~.  \eqno(7)
$$
With a tritium efficiency of of $\eta=0.5$, we therefore find that 400 t of DU correspond to 9.6 kg of tritium, which are equivalent to an explosive power of 0.5~$\times$~9.6~kg~$\times$~133~kt/kg~=~640~kt of high explosives.

\section{Delayed radiological effects of pure-fusion \\fourth-generation nuclear weapons}

The delayed radiological effects of the neutrons produced during thermonuclear explosions have been extensively discussed during the hydrogen bomb debate of the 1950s \cite{LEIPUN1957-}, and to a lesser extent during the neutron bombe debate of the 1980s \cite{SANDM1972-}. In the realm of non-military applications of nuclear energy, the problem of shielding nuclear reactors \cite{ETHERI1958-}, and particle accelerators \cite{HOYER1968-}, from the environment has lead to the study of similar problems.  Consequently, we know from these studies that there are only a few critical nuclear reactions, i.e., reactions producing radioactive nuclei with a life-time larger than a few years.

For neutrons escaping into the air, the critical reaction is the $^{14}N(n,p)^{14}C$ process in which carbon-14 is produced; and in interactions with the ground (e.g., the soil and concrete structures) the critical reactions are $^{23}Na(n,2n)^{22}Na$ and $^{59}Co(n,\gamma)^{60}Co$ in which sodium-22 and cobalt-60 are produced.  The radiobiological characteristics of these nuclei are listed in Table~1.  For interactions with other materials, e.g., iron, most activation products are short-lived and their effect can be considered as part of the prompt lethality of the weapon.

In order to simplify the discussion, we will further assume that the fourth-generation nuclear weapons under consideration can be considered of the ``pure-fusion'' type \cite{GSPON2000-}.   Therefore, we assume that they derive the bulk of their explosive power from the deuterium-tritium reaction, and that their overall radiological impact is dominated by this reaction so that the effect of other radioactive materials that may be present in the device can be neglected. 

\subsection{Neutron interactions with the air}

In good approximation, the vast majority of neutrons released into the atmosphere are ultimately captured by nitrogen which then decays into carbon-14.  Since each tritium nucleus fusing with a deuterium nucleus produces one neutron, $m_T$ g of tritium, $A=3$, will produce $\tfrac{14}{3} m_T$ g of carbon-14, $A=14$.  Therefore, the radiological burden due to neutrons interacting with air will be
$$
          B_a(^{14}C) =  \tfrac{14}{3} \phi_a \beta_{C} \eta m_T \eqno(8)
$$
where $\phi_a$ is the fraction of neutrons escaping into the atmosphere, and $\beta_C$ the specific burden of carbon-14 (see Table~1).

\subsection{Neutron interactions with concrete structures}

We assume for concrete a typical composition of the kind used in normal construction which contains about 0.5 \% of sodium per weight  \cite[Table.22a]{ETHERI1958-}.  For comparison, a special barytes concrete of the kind used in nuclear reactor shielding contains only about 0.1 \% of sodium per weight  \cite[Table.22b]{ETHERI1958-}.

The fraction of neutrons which interact with the sodium nuclei and produce the radioactive isotope sodium-22 is given by the macroscopic activation cross-section
$$
               \Sigma_{act} = \kappa n \sigma        \eqno(9)
$$
where $\kappa$ is the concentration of sodium nuclei in the concrete, $n$ the nuclear density, and $\sigma$ the fusion-neutron sodium-22 activation cross-section given by \cite{GARG-1977-} (see Table~2).

Similarly to the argument that lead to equation (7), since each tritium nucleus fusing with a deuterium nucleus produces one neutron, the radiological burden due to neutrons interacting with concrete is
$$
 B_c(^{22}Na) =  \tfrac{22}{3} \phi_c \frac{\Sigma_{act}}{\Sigma_{abs}} \beta_{Na} \eta m_T \eqno(10)
$$
where $\phi_c$ is the fraction of neutrons escaping into concrete and $\Sigma_{abs}$ the macroscopic fusion-neutron absorption cross-section of concrete \cite[p.186]{SEIFR1984-} (see Table~2).

\subsection{Neutron interactions with the soil}

Contrary to air, the chemical composition of the soil can considerably vary from place to place.  The average concentration of sodium in rocks can vary from less than 1 \% (average limestone) to about 4 \% (average igneous rock).  Moreover, the sodium concentration will obviously be significantly higher in coastal regions because of the proximity of sea water. In the case of cobalt, reference \cite{LEIPUN1957-} gives a concentration of $10^{-5}$ which seems rather high compared to other sources.  In the present paper we take the soil composition used in reference \cite{HOYER1968-}, a cobalt concentration of $10^{-6}$ together with the cobalt-60 activation cross-section given by \cite{LEIPUN1957-}, and the macroscopic fusion-neutron absorption cross-section in soil published in \cite[p.184]{SEIFR1984-} (see Table 3).

By the same reasoning that lead to equation (10), the contributions $B_s(^{22}Na)$ and $B_s(^{60}Co)$ to the radiological burden due to neutrons interacting with the soil are given by equations identical to (9) and (10), where all variables and parameters are given the appropriate values.

\section{Fourth-generation nuclear weapons yield equivalent do the radiological burden due to the use of depleted-uranium munitions}

Collecting the results of the two previous sections, the overall burden is
$$
  B(\eta,m_T) = B(^3T) + B_a(^{14}C) + B_c(^{22}Na)+ B_s(^{22}Na) + B_s(^{60}Co)  \eqno(10)
$$
which becomes, putting all the numbers in,
$$
  B(\eta,m_T) =   2 \times 10^6 (1 - \eta) m_T  
$$
$$
             +  \Bigl( 6.5 \times 10^3 \phi_a +
                  8 \times 10^4 \phi_c  +
                  6 \times 10^4 \phi_s  +
                  9 \times 10^3 \phi_s  \Bigr) \eta m_T    \eqno(11)
$$
where $\phi_a + \phi_c + \phi_s = 1$.  This expression gives, as a function of the tritium burn-up $\eta$, the radiological burden in units of rem due to the combustion of $\eta m_T$ g of tritium, which correspond to an explosive yield of
$$
                   Y = 133 ~ \eta m_T  ~~~ ~~~ \text{[t HE]}        \eqno(12)
$$
tons equivalent of high-explosives (HE).

Expression (11) shows that for ``ground bursts'' (i.e., explosions due to the engagement of targets on the ground) the radiological burden due to neutron interactions is dominated by the activation of concrete structures and the soil because in this case it is likely that $\phi_a < \phi_c + \phi_s$.  Moreover, the combined numerical factor obtained by adding the contributions of the activation of $^{22}Na$ and $^{60}Co$ in the soil is $6.9 \times 10^4$, i.e., nearly equal to the factor multiplying  the contribution of concrete.  Therefore, in the case of ground bursts, we can simplify (11) and write
$$
  B_g(\eta,m_T) \approx  
                         2 \times 10^6 (1 - \eta) m_T  + 
                         8 \times 10^4 \phi_g  \eta m_T       \eqno(13)
$$
where  $\phi_g=\phi_c+\phi_s$ is the fraction of neutrons interacting with the ground (i.e., not escaping into air).

The first conclusion that can be draw from (11) and (13) is that for tritium burn-ups on the order of $\eta \approx 0.5$ the radiological burden is to a good approximation determined by the dispersal of the unburnt tritium --- so that the assumption made in section 2 was justified.

The second conclusion is that in ground bursts the contribution from the ground materials's activation becomes only important when $\eta$ gets close to one, more precisely, when
$$
         \eta \geq 1 - \frac{8 \times 10^4}{2 \times 10^6} \phi_g  ~~,   \eqno(14)             
$$
i.e., $\eta \geq 0.96$ for $\phi_g=1$.

We can therefore make two extreme assumptions to estimate the amount of tritium, and thus the explosive yield, equivalent to the radiological burden of a given amount of DU:

\begin{itemize}

\item $\eta <  0.9$ ~~ Combining (7) and (12):
$$
       Y \approx 133 \frac{\eta}{1 - \eta} \frac{\beta_U}{\beta_T} m_U
         ~~~ ~~~ \text{[t-HE/g-DU]} ~~. \eqno(15)
$$ 

\item $\eta > 0.9$ ~~ Combining (3), (12), and (13):
$$
       Y \approx 133 \frac{1}{\phi_g} \frac{\beta_U}{8 \times 10^4} m_U
         ~~~ ~~~ \text{[t-HE/g-DU]}  ~~. \eqno(16)
$$ 
\end{itemize}

As can be seen, (15) corresponds to a lower-limit of the yield, while (16) gives an upper limit which corresponds to the fact that in high-tritium-efficiency pure-fusion weapons the radiological burden is dominated by the neutron activation of the materials in the proximity of the explosion.

\section{Conclusion}

The main technical result of this paper are equations (15) and (16) which give reasonable estimates of the total explosive yield of a hypothetical engagement of fourth-generation nuclear weapons that would result in the same radiological burden as a given total amount of depleted uranium expended in combat.  In deriving these equations the respective radiological burdens of the two types of weapon systems are used only as a means for comparing their relative medium- to long-term radiological impacts on human beings and the environment.  The absolute values of the burdens calculated in this paper should therefore not be trusted.  However, since similar conservative assumptions are made for both types of weapon systems, the \emph{ratio} of these burdens provides a significative means for comparing their respective delayed radiological impacts, and to derive the yields given by equations (15) and (16).

In particular, the assumption made in section 2 of ``total dispersal'' of the depleted-uranium is matched by the assumption that the unburnt tritium is also totally dispersed in the environment.  In order to improve the accuracy of our comparison one would have to study complicated scenarios of the engagment of both types of weapon systems, which may imply many self-compensating hypothesis that are not likely to significantly change our major conclusion --- \emph{that the expenditure of many tons of DU has a radiological impact comparable to the combat use of many kilotons of pure-fusion thermonuclear explosives}. 

Indeed, assuming a tritium burn-up of 50 \% ($\eta = 0.5)$ in equation (15), and that half of the neutrons interact with the ground ($\phi_g = 0.5)$ in equation (16), we get a total yield of circa 60 kt, respectively 3 Mt, in the expenditure of 40 t of depleted-uranium as in Yugoslavia --- and ten times more in Iraq.

These total explosive yields mean that in order to match the radiological burden due to the combat use of depleted-uranium in these countries, one could have hypothetically used several thousands precision guided delivery systems, each carrying a fourth-generation nuclear warhead with a yield in range of one to hundred \emph{tons} of high-explosive equivalent, instead of the few tens or hundreds of \emph{kilograms} of high-explosives currently delivered by these systems.

Such a possibility, which is far from being hypothetical \cite{GSPON2000-}, is shedding a dramatic light on the strategic-political significance of the breaking of the taboo against the battle-field use of depleted-uranium.  Of course, it has been repeatedly stated that the absolute burden due to the combat use of depleted uranium is small \cite{FETTE2001-,BEACH2001-}, but such a use is a violation of the spirit, if not the letter, of a norm that was in force since 1945.

\newpage


\begin{table}
\hskip 0.0cm
\begin{center} 
\begin{tabular}{|c|c|c|c|c|}
\hline
            &                          &                     &                   &                  \\
            &       {\bf half}         &  {\bf specific}     &     {\bf EDCO}    & {\bf specific}   \\
            &       {\bf life}         &  {\bf activity}     &                   &  {\bf burden}    \\
 units      &          year            &       Ci/g          &   rem/$\mu$Ci     &       rem/g      \\
 symbol     &       $t_{1/2}$          &     $\alpha$        &       $D$         &    $\beta$       \\
\hline 
{\bf T-3}   &           12.3           &  $1\times 10^{4}$   & $2\times 10^{-4}$ & $2\times 10^6$   \\
\hline
{\bf C-14}  &           5570           &        4.7          & $3\times 10^{-4}$ & $1.4\times 10^3$ \\
\hline
{\bf Na-22} &            2.6           & $6.4\times 10^{3}$  & $2\times 10^{-2}$ & $1.3\times 10^8$ \\
\hline
{\bf Co-60} &            5.26          & $1.2\times 10^{3}$  & $4\times 10^{-3}$ & $4.6\times 10^6$ \\
\hline
{\bf U-238} &     $4.5\times 10^9$     & $3.4\times 10^{-7}$ &          70       &         24       \\
\hline

\end{tabular}
\end{center}
\caption{Radiobiological characteristics of critical materials}  
\end{table}

\begin{table}
\hskip 0.0cm 
\begin{center} 
\begin{tabular}{|c|c|c|c|c|}
\hline
                          &                     &                    &                  \\
                          &     {\bf symbol}    &        {\bf   }    &    {\bf units}   \\
                          &                     &                    &                  \\
\hline 
{\bf mass density}        &      $\rho$         &        2.37        &      g/cm$^3$    \\
\hline
{\bf mean atomic weight}  &      $A$            &        25.7        &                  \\
\hline
{\bf nuclear density}     &      $n$            & $5.5\times 10^{22}$ &     cm$^{-3}$     \\
\hline
{\bf neutron absorption cross-section} & $\Sigma$  &      0.026         &       cm$^2$/g   \\
\hline

{\bf Na-23 concentration} &    $\kappa$         & $4.9\times 10^{-3}$ &  g-Na/g-concrete \\
\hline
{\bf Na-22 activation cross-section} & $\sigma$ & $8\times 10^{-27}$ &  cm$^2$          \\
\hline

\end{tabular}
\end{center}
\caption{Mean nuclear characteristics of Portland concrete } 
\end{table}

\begin{table}
\hskip 0.0cm 
\begin{center} 
\begin{tabular}{|c|c|c|c|c|}
\hline
                          &                     &                    &                  \\
                          &     {\bf symbol}    &        {\bf   }    &    {\bf units}   \\
                          &                     &                    &                  \\
\hline 
{\bf mass density}        &      $\rho$         &    $\sim$ 1.8        &      g/cm$^3$    \\
\hline
{\bf mean atomic weight}  &      $A$            &        21.7        &                  \\
\hline
{\bf nuclear density}     &      $n$            & $5\times 10^{22}$ &     cm$^{-3}$     \\
\hline
{\bf neutron absorption cross-section} & $\Sigma$ &       0.030         &       cm$^2$/g   \\
\hline

{\bf Na-23 concentration} &    $\kappa$         & $4.7\times 10^{-3}$ &  g-Na/g-soil \\
\hline
{\bf Na-22 activation cross-section} & $\sigma$ & $8\times 10^{-27}$ &  cm$^2$          \\
\hline
{\bf Co-59 concentration} &    $\kappa$         & $1\times 10^{-6}$  &  g-Co/g-soil \\
\hline
{\bf Co-60 activation cross-section} & $\sigma$ & $6\times 10^{-23}$ &  cm$^2$          \\
\hline

\end{tabular}
\end{center}
\caption{Average nuclear characteristics of dry top soil } 
\end{table}

\end{document}